# Numerical simulation of information recovery in quantum computers.

P. J. Salas

Depto. Tecnologías Especiales Aplicadas a la Telecomunicación,

E.T.S.I. Telecomunicación, U.P.M., Ciudad Universitaria s/n, 28040 Madrid (Spain)

A.L. Sanz

Depto. Física Aplicada a las Tecnologías de la Información,

E.T.S.I. Telecomunicación, U.P.M., Ciudad Universitaria s/n, 28040 Madrid (Spain)

PACS NUMBER: 03.67.Lx, 03.65Yz

# ABSTRACT

Decoherence is the main problem to be solved before quantum computers can be built. To control decoherence, it is possible to use error correction methods, but these methods are themselves noisy quantum computation processes. In this work we study the ability of Steane's and Shor's fault-tolerant recovering methods, as well a modification of Steane's ancilla network, to correct errors in qubits. We test a way to measure correctly ancilla's fidelity for these methods, and state the possibility of carrying out an effective error correction through a noisy quantum channel, even using noisy error correction methods.

# I. INTRODUCTION

Classical computers are nowadays an essential tool. The main idea supporting them is a classical logic implemented through a quantum hardware. Their reliability comes from the high developed technology used to implement logical gates and from their robustness against noise.

In classical theory, information does not depend on the physical system used to implement bits. A new way leading to quantum logic and quantum computation appeared once the physical nature of information was recognized.

The power of quantum computers stands on two strongly nonclassical features: interference and parallelism. Both of them require creation and manipulation of entanglement states involving large ensembles of quantum bits (qubits).

One of the fundamental obstacles in the development of quantum computers is the loss of coherence of the quantum states involved. Decoherence produces errors that have to be eliminated if progress in computation is desired [1]. There are several serious problems arising from the following facts:

1.- In contrast to classical errors, quantum ones are continuous, because they are related to the infinite number of coefficients used to describe the qubit.

2.- The impossibility (as a result of non-cloning theorem [2]) to use the classical copying information method in order to get enough redundant information encoded. Such a method would allow us to recover the information if an error occurred in the codeword.

3.- It is not (in general) possible to use the majority-voting method to read the information encoded in the computer's state, because this reading would collapse the state with an irreversible lost of information.

All these facts have slowed down the first steps in this field until Shor [3] and Steane [4] discovered the subtle ideas of quantum error correction (QEC) methods [5]. These methods allow us to correct errors in computer states without destroying their coherence, avoiding thus their collapse.

The first step is to transform continuum errors in the information-qubit (IQ in the following) into discrete ones [6]. This is not difficult if we realise that a general interaction with the environment (taking place when the initial IQ state $|\phi(0)\rangle = a|0\rangle + b|1\rangle$ decoheres by becoming entangled with the states of the environment) can be expressed as a linear combination of four possibilities, including three kinds of error:

$$|\phi(t)_{\text{corrupted}}\rangle =$$
$$(a|0\rangle + b|1\rangle) \otimes |e_I\rangle + \quad \rightarrow \text{No error}$$
$$(a|1\rangle + b|0\rangle) \otimes |e_x\rangle + \quad \rightarrow \text{Bit flip error}$$
$$(a|0\rangle - b|1\rangle) \otimes |e_z\rangle + \quad \rightarrow \text{Phase error}$$
$$(a|1\rangle - b|0\rangle) \otimes |e_y\rangle + \quad \rightarrow \text{Bit and phase error}$$

where $\{|e_i\rangle, i = I, x, y, z\}$ are the environment states (neither normalised nor orthogonal). Note that these states are related to Pauli matrices (indicated by the subindex), allowing us to interpret the evolution as a qubit having one of these four errors, even though this identification will only be strict when the orthogonal environment basis is used. To discern among the errors, an ancilla (a three-bit string is needed, at least) of orthogonal states is introduced to store the error information syndrome. At this point it is possible, by doing some collective measurement on the system, to work out the error. Afterwards, by inverting the transformation and disentangling the system from the environment it is possible to correct the qubit error.

In doing so, we have learned nothing about the coefficients a and b (because the measurement is collective), i.e. nothing about the quantum information encoded in the IQ. To fulfil this condition, each qubit of information is spread (encoded) in more than one qubit, in order to introduce some kind of redundancy, similar to the classical case and an adequate ancilla state has to be prepared. The encoding will be a unitary mapping of $H_2^{(1)}$ into $H_2^{(n)}$ ($H_2^{(1)}$ being the usual two-dimensional Hilbert space, and n the number of qubits used to encode).

A general IQ state $|\phi(0)\rangle$ is encoded using the logical $|0_L\rangle$ and $|1_L\rangle$ as $|\phi(0)_E\rangle = a|0_L\rangle + b|1_L\rangle$. This qubit could be affected by an error during the computation or transmission process, and if so, it must be corrected. In order to correct the three kinds of error considered previously, an ancilla state has to be prepared and, by letting it interact with the IQ, one can find a way to an information recovering method (R).

As the recovering process is a quantum one, it is strictly necessary to take into account the possible errors introduced by itself. The whole recovery network as well as the individual gates, must be implemented *fault tolerantly* [7] to control error spreading. We want to copy onto our ancilla the information about errors in the IQ block, without contaminating the IQ too much with phase or bit-flip errors, and without destroying its coherence. This property is fulfilled by the methods proposed by Steane and Shor, which we will review briefly in the following section.

In this paper, we examine the ability of a quantum error correction code to control decoherence of a qubit interacting with an environment. In spite of the fact that the correction procedure is itself faulty, we show that the whole process is advantageous.

The main steps of error recovery following Steane's and Shor's fault-tolerant methods are described in Sec. II. In Sec. III we compile the simulation results obtained for a recovery procedure in which errors follow an stochastic error model (described in Sec. III A). In Sec. III B we compare the fidelities of several ancilla's networks vs. different gate and evolution errors. Error spreading and fault-tolerant syndrome extraction in these networks are described in Sec. III C. Finally, in Sec. III D we put all those steps together at work in order to achieve an effective information qubit recovery.

## II. FAULT-TOLERANT QUBIT RECOVERY

Suppose the initial IQ state $|\phi(0)\rangle$ is involved in a quantum computation or transmission processes. Because of decoherence, the qubit will be corrupted and our task is to correct it in order to recover as much information as possible. The whole recovery method must be carried out fault tolerantly, and we can recognise the following general steps:

### A. Encoding qubits.

Calderbank, Shor [8] and Steane [9] (CSS codes) have proposed a method for achieving the criteria of fault tolerance. The simplest example of the encoding protocol is a generalisation of the classic linear 7-bit Hamming code. In this code, [7,4,3], a 7-bit block is used to encode 4 bits of information ($2^4$ strings) and a 3 bit syndrome ($2^3$ strings) to store the error information (or syndrome). In the quantum case, $|0\rangle$ and $|1\rangle$ qubits are encoded, not directly with codewords as in classical encoding, but with *cosets*. The starting point is a subcode $C^\perp \equiv [7,3,4]$ (dual of the Hamming code $C \equiv [7,4,3]$) and the encoding uses the $C^\perp$ *cosets relative* to C. Specifically the logical $|0_L\rangle$ is encoded as the linear combination of all codewords from the $|0000000\rangle$ coset (eight even parity

Hamming codewords) and the logical $|1_L\rangle$ is encoded as the linear combination of all $|1111111\rangle$ coset codewords, having odd parity:

$$|0_L\rangle =$$
$$\frac{1}{\sqrt{8}}\left\{\begin{array}{l}|0000000\rangle+|0001111\rangle+|0110011\rangle+|0111100\rangle+\\|1010101\rangle+|1011010\rangle+|1100110\rangle+|1101001\rangle\end{array}\right\}$$

$$|1_L\rangle =$$
$$\frac{1}{\sqrt{8}}\left\{\begin{array}{l}|1111111\rangle+|1110000\rangle+|1001100\rangle+|1000011\rangle+\\|0101010\rangle+|0100101\rangle+|0011001\rangle+|0010110\rangle\end{array}\right\}$$

It is not difficult to construct the quantum network to implement this encoding, taking into account the code generation matrix.

### B. High fidelity ancilla synthesis.

A single error in the ancilla's synthesis could propagate into two or more errors (by means of CNOT gates) and could cause the IQ contamination, with no possibility for correction. So we need to synthesise high fidelity ancillas, checking them carefully before letting them interact with the IQ.

In this work we will study specifically two different ancillas defining two recovery methods: Steane's and Shor's.

#### *1. Steane's ancilla* [10]

To prepare the ancilla state, Steane's method uses the $C^\perp$ generator matrix $G_{C^\perp} \equiv H_C$ (C parity check matrix):

$$G_{C^\perp} = \begin{pmatrix} 0 & 0 & 0 & 1 & 1 & 1 & 1 \\ 0 & 1 & 1 & 0 & 0 & 1 & 1 \\ 1 & 0 & 1 & 0 & 1 & 0 & 1 \end{pmatrix}$$

This prepares a $|0_L\rangle$ state (see Fig. 1, Network-1, left side G) and, by doing a seven bitwise Hadamard rotation at the end, we would finally get an ancilla state $|\alpha_{Steane}\rangle = \{|0_L\rangle + |1_L\rangle\}/2^{1/2}$. The second part of the network, shown in Fig. 1 as V, detects a possible bit-flip error using an eighth verification qubit to store this possibility. If the measurement result of the eighth qubit value is 0, the ancilla could be correct, otherwise (if the eighth qubit value is 1), bit-flip error has been detected and the ancilla rejected. In this sense, this network is able to detect ancillas infected with a one bit-flip error.

By fault-tolerant we mean that if the evolution or gate error probability are ε and γ (respectively), the network's error probability behaves like $O(\varepsilon^2, \gamma^2)$. More generally, for a t-error correcting code, we would require the probability for the state to have an error of weight larger than t to be $O(\varepsilon^{t+1}, \gamma^{t+1})$. The ancilla's Network-1 is not fault tolerant against phase-errors because CNOT gates propagate errors in the G step, but these errors are not so harmful (as we will see in paragraph III C, "Syndrome extraction and measurement"). On the contrary, Network-1 is fault-tolerant against bit-flip errors, if the verification step (V) is included (re-starting the network whenever the eighth verification qubit is found to be 1) and the whole IQ recovery is considered (see next Fig. 8).

To verify that Network-1 is indeed fault tolerant against bit-flip errors when the verification is included, we divide it in two pieces: the initial $|0_L\rangle$ generation G and the subsequent verification V. Whenever no failure occurs during V (correct), the final $|0_L\rangle$ state is guaranteed to be free of bit-flip errors. If G is correct too (case 1 in Fig. 2), all the process is correct. In case 3 of Fig. 2, V is correct but G fails. If we consider the $|0_L\rangle$ as the linear combination:

$$|0_L\rangle = \frac{1}{\sqrt{8}} \sum_{x \in C^\perp} |x\rangle$$

and $X_e$ is a bit-flip error of weight $w = W(e)$, then $X_e|0_L\rangle$ will be a linear combination of kets such as $|x+e\rangle \in e+C^\perp$, and all of them are in the same coset of $C^\perp$ with respect to the total Hilbert space $H_2^{(7)}$. As we use $C^\perp \equiv [7,3,4]$, with a four-bit syndrome, there is a one-to-one correspondence between the 16 cosets and their syndromes. In this code, all errors with weight $w \geq 4$, are equivalent to errors with $w \leq 3$ (corresponding to 16 different errors) because whenever an error e, with $W(e)>0$, happens in the $|0_L\rangle$ state, the effective error weight with respect to $|0_L\rangle$, is given by:

$$W_{eff} = \underset{V \in C^\perp}{Min} \{W(V \oplus e)\}$$

After measuring the syndrome, it is possible to recover the $|0_L\rangle$ ancilla state completely. So, if V step is correct, the ancilla will have no bit-flip error (of any weight).

Failure occurring during V but not in G (case 2 in Fig. 2), can result in bit-flip errors of all weights $w$ (all these are equivalent to $w \leq 3$ errors), but as they are uncorrelated, the bit-flip error probability will be $O(\gamma^w, \varepsilon^w)$. In this case the non-detected bit-flip errors at the end of the $|0_L\rangle$ Network-1 (bangs appearing in Fig. 1), are transformed into phase errors after the final Hadamard rotation to get $|\alpha_{Steane}\rangle$. Finally, we must account for the case when both G and V contain failures (case 4 in Fig.2). This can result in errors of any weight in the final state, but as they are uncorrelated, it is itself a second order process, and has $O(\gamma^2, \varepsilon^2)$ probability.

Neither any number of phase-flips errors are detected in G or V steps, nor one bit-flip error appearing at the end (bangs in Fig. 1) of the V network. These errors ($O(\gamma,\varepsilon)$) will be transformed (after the final bitwise Hadamard gates) to get $|\alpha_{Steane}\rangle$ into bit-flip errors in the former case, and phase-flip errors in the latter. Fortunately, some of these errors can be detected in the whole syndrome's qubit repetition (as we will see in the following, Fig. 10).

### 2. Shor's ancilla

Shor's method [11] proposes to measure the syndrome with a different ancilla state, and using a different strategy. The ancilla is prepared starting from a 4-qubit "cat" state (because four checks are needed to obtain each bit of syndrome):

$$|cat\rangle = \frac{1}{\sqrt{2}}\left(|0000\rangle + |1111\rangle\right)$$

which is four times Hadamard rotated to get Shor's ancilla state:

$$|\alpha_{Shor}\rangle = \frac{1}{\sqrt{8}}\begin{pmatrix} |0000\rangle + |0011\rangle + |0101\rangle + |0110\rangle \\ + |1001\rangle + |1010\rangle + |1100\rangle + |1111\rangle \end{pmatrix}$$

Shor's method uses a simpler ancilla state $|\alpha_{Shor}\rangle$ than Steane's one. The network needed to synthesise it is detailed in Fig. 3. This network introduces a fifth qubit to verify the possible bit-flip error. This kind of error will be converted into a phase flip error (as in Steane's network) through bitwise Hadamard gates at the end, and will spread it back to the IQ $|\phi_E\rangle$, damaging the information. To control these errors one has to notice that for all the ways a single gate may produce two bit-flip errors in the cat-state, the first and fourth qubit have different values. Using two CNOT gates with a fifth qubit as a target and by measuring it, we are able to detect these errors. The ancilla state is rejected if the result of the measurement is "1". In spite of all that, some errors could slip through this verification if they happen at the end of the network, after the last CNOT gate, but they are uncorrelated.

Therefore, we can conclude that the ancilla's synthesis (in both cases) can be done fault-tolerantly only against bit-flip errors, i.e. with an unrecoverable error probability behaving as $O(\varepsilon^2, \gamma^2)$.

## C. Syndrome extraction.

In this process the error must be copied onto the ancilla state, neither introducing unrecoverable errors ($O(\varepsilon^2, \gamma^2)$) nor destroying the IQ coherence. The first condition is fulfilled by means of CNOT and H gates implemented fault-tolerantly, applying them transversally [12]. The second requirement is already taken into account in the special ancilla state used.

The ancilla state interacts with the IQ to extract the syndrome, following the general process:

$$S\{|\alpha\rangle \otimes |\phi(t)_{E,Corrupted}\rangle\} = |\alpha_S\rangle \otimes |\phi(t)_{E,Corrupted}\rangle$$

For $|\alpha\rangle$, we use Steane's ancilla state or Shor's. Now the syndrome s has been moved to the ancilla state $|\alpha_S\rangle$.

To get the syndrome, Steane's network applies the H and CNOT gates in which each qubit from a codeword interacts at most with one qubit of another codeword. The IQ syndrome is accumulated into the ancilla. Specific quantum networks based on the [[7,1,3]] code will be detailed in the following to carry out this syndrome extraction. In Shor's method the ancilla's parity is measured. If the IQ is correct, $|\alpha_{Shor}\rangle$ has even parity, otherwise has odd parity. Notice the ancilla measurement is not an essential ingredient but simplifies the whole recovery method

Errors occurring in the middle of the IQ recovering process or coming from an erroneous ancilla synthesis have a special importance. These $O(\varepsilon,\gamma)$ errors, could contaminate both the IQ (propagated forward or backward by the CNOT ancilla-IQ gates) and produce a wrong syndrome. To eliminate this possibility, the syndrome will be repeated several times, choosing the correction action by a majority vote. Syndrome repetition makes the wrong syndrome probability $O(\varepsilon^2,\gamma^2)$ and the whole method fault-tolerant.

### D. Syndrome measurement.

By measuring the ancilla state we know the syndrome, and this enables us to identify the IQ error. To measure the bit-flip error, the parity check matrix is used in the computational basis $\{|0\rangle,|1\rangle\}$, and to detect a phase-flip error, a change of basis (bitwise Hadamard rotation) to $\{(|0\rangle+|1\rangle)/2^{1/2},(|0\rangle-|1\rangle)/2^{1/2}\}$ has to be carried out. With this change of basis, the phase-flips are transformed into bit-flips, and once they have been corrected, a new Hadamard rotation gets back to the computational basis.

From the code's point of view, the measurement of the syndrome is carried out by multiplying (mod 2) the parity check matrix $H_C$ times any codeword of the linear combination. Suppose the correct codeword is u, and the error is represented by the vector $e_i = (0 \ldots 1^{(i)} \ldots 0)$, having a 1 in the i-th position. A noisy codeword would be $u+e_i$. Then, by applying $H_C (u+e_i) = H_C e_i$, the result will be equal to one of the columns of $H_C$ matrix. Once the syndrome is known, it will be possible to correct the corrupted codeword to get the correct one: u. The [[7,1,3]] quantum code, including two quantum codewords, being able to correct one error (bit-flip, phase-flip or both) has been built starting from the classical Hamming [7,4,3] code. Unfortunately, for the codes we use, this works only in the single error special case. As we will see next, some errors are neither detected nor corrected.

Remark that, once we get the syndrome, it is not affected by errors, because it actually is a classical information and can be stored for a long time.

### E. Information-qubit correcting method.

When the error has been identified by means of the right syndrome, we proceed to correct it in the IQ, applying the inverse transformation. The whole recovering process can be summarised as:

$$R\{|\alpha\rangle \otimes |\phi(t)_{E,Corrupted}\rangle\} = |\alpha_S\rangle \otimes |\phi(t)_{E,Corrected}\rangle$$

At the end of this recovery method the $|\phi(t)_{Corrected}\rangle$ is obtained and we have to test how good our whole correcting process is. One way to quantify it is by using the *fidelity* concept, given by the overlapping between the final state and the error free state, i.e., the probability of obtaining the correct state vector once the error correction has been carried out:

$$F = |\langle \phi(t)_{E,Corrected} | \phi(0)_E \rangle|^2$$

In a general process, the error free initial IQ state vector $|\phi(0)\rangle$ (or its encoded state) is not known and the fidelity definition must be more precise, but for the present simulation we are able to prepare the error free IQ state (with a=b), to introduce an error on it and, finally, try to recover the original state. In this sense, it is possible to calculate the fidelity with the latter equation. Strictly speaking, the fidelity F = F(a,b) is defined as its minimum value when a and b change. The quantum code must maximise this minimum.

## III SIMULATION RESULTS

### A. Error model

To simulate the behaviour of quantum networks we use an *independent stochastic error model* based on the notion of error locations [13]. These are chosen as the locations in the network where errors take place, and have to satisfy some properties. At each location the error is introduced at random and independently of other errors in the same or in different locations. We consider that all the quantum steps have some error probability, and distinguish between *memory errors* (due to qubits evolution) with error probability $\varepsilon$, *one qubit gate errors*, due to one qubit gate operations (like measurements or Hadamard gates) with $\gamma$ error probability and *two qubit gates* as CNOT, with an error probability proportional to $\gamma$.

Memory errors are located at each time step in the network, affecting all the qubits evolving in that step. Their effect is highly related to the degree of parallelization in the network. One way to reduce this error is to increase the network parallelism. Following this idea we will introduce a new and improved network in this work (see Fig. 7, Network-3). To model the evolution errors we consider the depolarising channel model. For each error location affecting one qubit with an ε error, we consider an isotropic ε/3 error probability for the $\sigma_x$, $\sigma_y$ and $\sigma_z$ cases.

For the noisy one-qubit gates (Hadamard and measurement), the γ probability is introduced at each gate location. In the two-qubit gates (CNOT) case, we assume there are sixteen possibilities corresponding to the tensor product {I, $\sigma_x$, $\sigma_y$, $\sigma_z$} ⊗ {I, $\sigma_x$, $\sigma_y$, $\sigma_z$}. If the one qubit gate error probability is γ, each two-qubits error appears with probability γ/15, because the I⊗I term is not, actually, an error operation. To introduce the noise, we let the gate operate before the error is introduced. This O(γ) (instead of O($\gamma^2$)) two-qubit error behaviour clearly over-estimates the difficulty of error correction, although it is not an unrealistic assumption.

We have to point out that neither leakage error nor explicit assumptions on scaling problems are taken into account. We only assume that errors ε and γ are independent of the total number of network qubits. To introduce errors in this way is equivalent to collapse stochastically the $|\phi(t)_{Corrupted}\rangle$ qubit state in one of the correctable error terms. To choose the kind of error, a Monte Carlo simulation is invoked by means of a random generator number program based on the Prime Modulus Multiplicative Linear Congruential Generator. The form of the multiplicative congruential generators is:

$$X_i = C \, X_{i-1} \, \mathrm{mod}(2^{31} - 1)$$

Each $x_i$ is then scaled into the unit interval (0,1). If the multiplier C is a primitive root modulo $2^{31}$-1 (which is prime), then the generator will have the maximal period $2^{31}$-2. The faster choice is C=16807.

We have checked the convergence of our results in several ways, testing the dependence of $|0_L\rangle$ fidelity preparation with Network-1 on several parameters:

a) We have repeated the simulations several times with the same C value to verify the result's spreading. It has been carried out using a number of steps about ten times the inverse of the error probability. The spread is never bigger than 1% referred to the average value.

b) Extending the number of steps to 1000 times the inverse of the error probability, the results were always within the error bars obtained in the previous case.

c) Several other multiplier values: C=397204094, and C=950706376 including shuffling have been used. In no case is the variation of our results bigger than 1%, referred to the average value.

## B. Ancilla's synthesis simulation

In the numerical simulation of the ancilla's preparation, we pay a special attention to bit-flip errors eliminated through a special ancilla verification network. Steane's method takes advantage of the fact that the encoded state corresponds to a linear combination of [7,3,4] classical Hamming codewords. In this case it is possible to use a parity check matrix to detect single bit-flip errors. This kind of error is extremely harmful because at the end of Network-1, Hadamard rotations convert them into phase-flip errors and they are able to spread back into the IQ, damaging it.

The phase-flip errors are not so awful, because Hadamard gates (at the end of the network) will convert them into bit-flip errors that could be detected in the (possible) syndrome repetition process.

To check the validity of this ancilla's preparation, we calculate its fidelity $F_a(\varepsilon,\gamma)$ as previously defined, and according to case 2 in Fig. 2 (and the fact that the network is not fault-tolerant against phase–flips) we expect the following behaviour:

$$F_a(\varepsilon,\gamma) \approx 1 - O(\varepsilon,\gamma)$$

In the *ancilla factory error correction* introduced by Steane [10], the most difficult step is to prepare, or synthesise, the ancilla state to put it in interaction later with the IQ system in order to correct the possible errors. As previously stated, we need two ancillas (see Fig.8): one to correct bit-flips ($a_x$, in the computational basis) and a second one to correct phase-flips ($a_z$ having changed to a rotated basis in the IQ).

First of all, to simulate the behaviour of Steane's ancilla synthesis, we prepare the $|0_L\rangle$ state using the G noisy piece of Network-1 and then correct the possible errors with the final verification (V) step. At the end of this process we will get the $|\alpha'_{Steane}\rangle = (|0'_L\rangle + |1'_L\rangle)/2^{1/2}$ (primes meaning some errors could infect the ancilla) $\approx (|0_L\rangle + |1_L\rangle)/2^{1/2} = |\alpha_{Steane}\rangle$ state. Then we calculate the ancilla's fidelity as the overlap $|\langle\alpha'_{Steane}|\alpha_{Steane}\rangle|^2$. In order to get a good statistic for the fidelity, we make a number of runs much bigger than the maximum value of {$1/\varepsilon$, $1/\gamma$} (minimum ten times bigger), to obtain the average value for the fidelity.

In Fig. 4(a) and 4(b) we compare the ancilla's fidelity (in fact, 1-$F_a(\varepsilon,\gamma)$) for three simulations corresponding to Network-1 both with and without perfect and noisy verification V. They also show a theoretical estimation for the fidelity of the ancilla synthesis without the verification step, based on the same considerations made in Steane's paper [10]. Taking into account the three kinds of error (because no verification step is used), it is possible to approach the fidelity as $(1-\gamma)^{19}(1-\varepsilon)^{72}$. Note that the behaviour of the (1-$F_a(\varepsilon,\gamma)$) is not as $O(\varepsilon^2, \gamma^2)$ because the network is not fully fault-tolerantly constructed. In Fig. 4(b), the gate error dependence of the ancilla's fidelity is shown, displaying a lower $\gamma/n$ (n is the number of encoding qubits, n=7) dependence than in the $\varepsilon$ case. To balance the values of both errors, we represent the fidelity vs. $\varepsilon$ and $\gamma/7$ (so they fulfil for approximately $7\varepsilon=\gamma$).

As we can see in Fig. 4(a), we get a better fidelity by using an ancilla with perfect verification. But surprisingly, ancilla's synthesis including the noisy verification network (V) is not extremely advantageous compared to the case when the verification is not used (undressed ancilla network), especially in small enough $\varepsilon$ and $\gamma$ regions. In the case of big $\gamma$ ($\approx 0.01$, Fig. 4(a)), noisy verification seems to have a small improvement than in the case of an undressed ancilla network for small enough $\varepsilon$ error. This may be understood if we realise that our verification error network tries to eliminate single bit-flip errors, but at the same time in the middle of this process, it also introduces some unrecoverable errors. Moreover, the errors that took place at the end of the recovering process are not detected, even in the case of only one bit-flip or phase error, because they are not reflected in the eighth qubit. Some of them are shown in Fig. 1 as bangs.

We reach similar conclusions for the $\gamma$ gate error dependence (Fig. 4(b)) when error $\varepsilon$ is fixed. A lesser $\gamma$ dependence is appreciated compared to $\varepsilon$ variable. For both $\varepsilon=0.01$, 0.001 the ancilla's fidelity including V step is the worse case, and smaller $\gamma$ values are needed to improve the results.

Instead of preparing a full ancilla state $|\alpha_{Steane}\rangle = (|0_L\rangle+|1_L\rangle)/2^{1/2}$ (theoretically needed), we are now going to prepare only a $|0_L\rangle$ qubit state, more closely related to the

full recovery network (see Fig. 8) as it is the common element appearing in both ancillas's synthesis. In this case we have eliminated the Hadamard gates at the end (Fig. 1), and we only care about the bit-flip errors. However, as has already been pointed out, bit-flip and phase-flip errors are not equally bad, but in both cases the fidelity could vanish. In order to distinguish this difference, it is worth to introduce the probability $P_{bf}(\varepsilon,\gamma)$ of two-bit-flip errors not detected in the ancilla synthesis. These errors will contaminate the IQ in an unrecoverable way, when the full recovery network (Fig. 8) is performed. This probability is calculated as the total number of simulations with two-bit-flip errors not detected divided by the total number of simulations.

In this case we reach similar conclusions on the general $\varepsilon$-$\gamma$ error behaviour as previously stated. In Fig. 5(a)-(d) we show the simulation results for the $|0_L\rangle$ preparation. Looking closely at Fig. 5(b) and 5(d), the probability for two-bit-flips not detected are very different in the cases when we use V step or not. This difference is not reflected in the fidelity (Fig. 5(a) and (c)). The $P_{bf}(\varepsilon,\gamma)$ seems a good way to distinguish the ancilla's quality in order to be included in the whole recovery IQ network.

Taking advantage of the fact that not all error propagation routes are open in the ancilla preparation network, Steane proposed another network [10] (called Network-2) with a smaller number of gates. His modification improves the network's parallelism slightly from 1.31 average gates/time step (for Network-1), to 1.35 gates/time step (Network-2). We have verified a small improvement in the fidelity using Network-2, but it is not shown in the figures.

In this work a new ancilla's network is proposed (see Fig. 7, Network-3). Four new qubits are introduced to accumulate the four parity checks instead of only one, as in Network-1 (the eighth qubit), allowing us to increase the parallelism in the $|0_L\rangle$ synthesis (1.31 gates/time step in Network-1 to 3.1 gates/time step), because now it is possible to perform several CNOT gates in parallel and to measure the syndrome in only one time step. Implementing the network with the appropriate technology does not seem a problem, for instance, with ion-trap methods. Remember that Network-3 only synthesises the state $|0_L\rangle$. If we want to have a $(|0_L\rangle+|1_L\rangle)/2^{1/2}$ state, we just need to add a final bitwise Hadamard rotation.

In Fig. 6 we show the simulation results obtained using Networks-1 and 3 as well as the fidelity for Shor's cat state synthesis. The reason for choosing the cat state and not the full ancilla state is analogous to the $|0_L\rangle$. Steane's state. In spite of Shor's network not being strictly comparable to Networks 1, 2 or 3, the simulation results of the cat state (Fig. 3) shows that it is more precise than Steane's $|0_L\rangle$. This behaviour is logical because the former network uses fewer time steps and gates than the latter. We observe that Network-3 has a lesser probability for two-bit-flip errors not detected (they will not be detectable by the recovery process) than Network-1. As Network-3 provides the best fidelity too, we may guess it will give the best results. The final comparison between the different methods will be carried out analysing the results for the full IQ recovery process.

### C. Syndrome extraction and measurement

The network used to correct the bit-flip and phase-flip errors (and both at the same time), takes into account the fact that we can measure the bit-flip syndrome just by applying a (transversal) CNOT gate between each IQ into the corresponding ancilla's qubit state. In the phase-flip error case, we previously rotate the IQ from the computational basis to the dual one (transforming the phase-flips into bit-flip errors), we apply the corresponding CNOT gates and get them back to the computational basis by applying Hadamard gates again.

In Steane's correction method, the whole ancilla-IQ network to measure the syndrome is shown in Fig. 8.

Each vertical dashed line corresponds to a whole time step. It is easy to understand how this network has been built if we realise, first of all, that we prepare an ancilla $a_z$ and copy the possible phase-flip error, using their back propagation target-to-control. Then we transform these phase errors in a qubit syndrome by rotating them. In the bit-flip error detection, the rotation has to be done before the CNOT gates in order to get the syndrome bits. In this recovering network, the gates are applied transversally, i.e. using encoded qubits, according to fault-tolerant ideas. Each gate affects the seven qubits simultaneously, and fourteen CNOT gates carry out the ancilla-IQ interaction.

However, in this whole recovery network there could be some non-detected errors coming from:

*Evolution IQ errors non-detected*, because they happen at the end or in the middle of the checking processes.

*IQ infections coming from the ancilla's interactions*. This happens when a CNOT gate connecting the IQ (as control) and the ancilla qubit (as target) is applied and a phase error happens in the ancilla qubit. The phase error is transmitted back to the IQ.

*Noisy gates*. The CNOT, Hadamard or measurement gates could fail themselves and introduce errors in both information and ancilla, and then propagate.

*Wrong ancilla synthesis*. Because of non-detected errors in the ancilla synthesis.

The worst errors are those happening in any location at the end of the recovery network, because they are neither detected nor corrected and would accumulate throughout the whole process.

Bearing in mind these considerations we are going to analyse the two syndrome extraction procedures referred in the general introduction.

Figure 9(a) shows two routes for the different kinds of possible non-detected errors producing a phase error in the IQ. Along the upper path, the phase errors in the $a_z$ synthesis, CNOT gates or evolution steps are converted in the H gates into bit-flip errors giving rise to a wrong syndrome, corrupting finally the information after the IQ correction. A bit-flip error introduced in the Hadamard gate itself would produce the same effect. On the other hand (lower path), bit-flip errors in the $a_x$ synthesis or in a free step evolution, are converted into phase errors by an H gate, infecting back the IQ through the CNOT gate. The Hadamard gate itself could introduce phase-flip errors. This contamination route coming from bit-flip errors in the ancilla synthesis is very harmful. It was for this reason we were so careful in trying to eliminate them using a verification step in the ancilla's network. The possible phase-errors ($O(\varepsilon,\gamma)$) in the IQ coming from ancilla infection can be detected in the next correction step.

Similarly, (Fig. 9(b)) bit-flip errors can be spread into the IQ. However, some of the possible bit-flip errors occurring in $a_z$ synthesis or in the subsequent evolution free step are not very harmful in spite of the CNOT gate connecting the ancilla and the IQ (soft arrow and cross). These possible bit-flips would be detected by the CNOT gates in the $a_x$ ancilla state and, finally, corrected. Some phase errors in the $a_x$ ancilla synthesis (not detected by the verification in Networks-1 or 3, for instance), are transformed into bit-flip errors in the H gate, and will provoke a faulty correction of the IQ because the wrong syndrome is used. The H gate and syndrome measurement themselves can introduce a bit-flip error that will also produce an erroneous syndrome.

Phase-flip errors in the ancilla's synthesis (upper path, Fig. 9(a) and the lower in Fig. 9(b)) will end up in a wrong syndrome. By repeating the syndrome and choosing for the final correction the majority-voting syndrome, we will get a greater probability of success in the IQ correction. In the simulation we calculate two syndromes: if they agree, the correction is carried out. Otherwise, a third one is calculated and we take the majority vote as the right syndrome. If the three syndromes are wrong (O($\varepsilon^3$, $\gamma^3$), we do nothing in that step. The probability for two of the three syndromes being equal and both wrong is smaller (O($\varepsilon^2$,$\gamma^2$)) than the probability of only one wrong syndrome. The whole recovery network for a three-syndrome repetition is detailed in Fig. 10.

The worse case will be when two or more bit-flip errors accumulate in the ancilla ($|0_L\rangle$) or in the IQ because we are not able to correct them. For this reason we have calculated the probability for two-bit-flip errors not detected in order to estimate the $|0_L\rangle$ quality in the ancilla's synthesis simulation (Fig. 5(b), (d) and Fig. 6(b), (d)). If a bit-flip error is detected, this ancilla state is rejected, starting the synthesis again. Having in mind this requirement, we assume the possibility of having some kind of device capable of preparing as many ancillas as we need at any time.

In Shor's method (for details, see [7]) we get the syndrome by carrying out two sets of CNOT gates according to the parity check matrix $\mathbf{H}_C$. In the computational basis we get the bit-flip syndrome and, in the rotated basis (after Hadamard gates) we get phase-flip syndrome, i.e. in total a six qubits syndrome. The general frame is similar to Steane's. To measure each qubit syndrome, a cat or ancilla state $|\alpha_{Shor}\rangle$ is needed. Each ancilla is a linear combination of even parity four qubit codewords. If the system has only one bit-flip error (for instance), the CNOT gates will change the parity in the ancilla codewords. Then by measuring this parity, the ancilla vector collapses in a particular codeword that provides the one bit syndrome. The lack of information about the codeword in which it collapses is related to the requirement of copying into the ancilla just information about the error, but not about the initial IQ.

Shor's ancilla synthesis is easier than in Steane's (or Network-3), but in the whole recovery IQ, the interaction ancilla-IQ takes 24 CNOT gates in the former, so long as the latter only needs 14 CNOT gates (for a one syndrome repetition). The whole Shor recovering network (measuring one syndrome) is similar to the one detailed in Fig. 10 for Steane's three-syndrome measurement method, but each CNOT gate now corresponds to a four one-qubit gate. In the case of three-syndrome repetition, to measure each syndrome, the whole recovery network would include 72 CNOT gates!. The same comments made before about the paths of errors infecting the IQ are valid for Sohr's method.

## D. Information-qubit correcting method

We have carried out the simulation for both methods (Steane and Shor) to recover the information for an encoded qubit (IQ) prepared in the $|\phi(0)_E\rangle = (|0_L\rangle + |1_L\rangle)/2^{1/2}$ initial state. In the IQ encoding, evolution, gate, phase and bit-flip errors are permitted. After the complete recovering, we calculate the quality of the IQ through the fidelity as the square overlapping between the error free state $|\phi(0)_E\rangle$ and the corrected state $|\phi(t)_{E,Corrected}\rangle$ (perhaps including some errors).

Figures 11(a) and 11(b) show the results for the IQ recovery fidelity vs. evolution ($\varepsilon$) and gate error ($\gamma/7$). In spite of Shor's cat state synthesis being easier than $|0_L\rangle$ Steane's state, in the case of the whole IQ recovery processes, the latter provides a greater fidelity than the former. This behaviour can be related to the fact that Steane uses 14 CNOT gates per total step syndrome correction so long as Shor uses 24 CNOT gates.

It is interesting to note that Shor's fidelity using one syndrome measurement is better than the value obtained by measuring three (for the interval of errors studied). This behaviour indicates that the interaction ancilla-IQ increases considerably from 24 to 72 CNOT gates (needed to measure three syndromes). A smaller error tax is needed to show the benefit of a three syndrome scheme.

Figure 11(a) and (b) also shows the IQ fidelity obtained using the ancilla's Network-3, proposed in this work. Their results represent a remarkable fidelity improvement in the total recovery processes, compared to the rest of networks studied. This is due to the fact that we use a more accurate ancilla synthesis method represented by Network-3, as we previously studied. Remember Network-3 was the one which had the smaller two-bit flip error probability (Fig. 6(b) and (d)).

The effect of including ancilla's verification step (V) in the IQ recovery process is shown in Fig. 12. The behaviour reflects (according to Figs. 6(b) and (d)) the adequateness of the criterion (probability for two-bit-flip errors not detected) chosen to quantify the quality of the ancilla. Specially in the small error region the inclusion of verification seems to be advantageous.

## IV CONCLUSIONS

We have carried out a classical simulation, i.e. using a classical computer, of a quantum error correction processes, with two different initial strategies: Steane's and Shor's. Evolution and gate errors reflecting decoherence and hardware errors, with different probabilities, are taken into account. We use a Monte Carlo method to introduce all three kinds of errors into some network locations, each one occurring with the same probability at every point of the network.

The simulation includes the ancilla synthesis, using several networks with an increasing parallelism. Because of the way in which ancilla's networks are arranged in the whole information-qubit recovery network, only bit-flip errors (if the H gates are not included at the end of the network) have to be controlled, because they are the most harmful. If some phase errors contaminate the ancilla, the syndrome repetition would control them. The worst errors are two or more bit-flips in the ancilla's sythesis, because they are not detected and are able to contaminate the information-qubit. In this sense, ancilla's fidelity does not constitute a good measurement to reflect its real accuracy. We introduce a better way to measure ancilla's capability for this recovery process: to calculate the non-detected two bit-flip error probability. If an ancilla verification is carried out, this probability will decrease. The ability of the non-detected two bit-flip error probability to measure the success of the whole recovery method is shown when a qubit is sent along a noisy quantum channel.

We propose a new Network-3 (Fig.7) for the ancilla's synthesis, providing, in all comparable cases studied, a good (low error probability) ancilla state. Shor's ancilla is always better than the previous one but it is not strictly comparable, because it involves a smaller number of gates. But, in the whole recovery network, this behaviour is inverted, and we get the best information-qubit recovery using Network-3 for the ancilla synthesis, repeating the syndrome and using a majority vote method (Fig.11). We confirm the benefits of using a verification step in the ancilla synthesis (Fig.12). Finally we conclude that Network-3 reveals to be the best (among those studied) ancilla's synthesis method (with syndrome repetition), for a full information-qubit recovery.

## ACKNOWLEDGEMENTS

We wish to thank Dr. A. Steane for his helpful discussions concerning the model. This work has been supported by Spanish Ministry of Science and Technology Project N. BFM2000-0013.

## FIGURE CAPTIONS

FIG. 1: Network-1: Steane's ancilla preparation. It synthesises the full $|\alpha_{Steane}\rangle = (|0_L\rangle + |1_L\rangle)/2^{1/2}$ state. The symbols are: (H) Hadamard, (O) is the $|0\rangle$ initial physical qubit, (M) measurement, (bangs) represent errors and the other symbols are CNOT gates. Each vertical dotted line corresponds to a time step.

FIG.2: Bit-flip error behaviour in the Steane's $|0_L\rangle$ synthesis.

FIG. 3: Shor's ancilla network.

FIG. 4: Difference to one of the ancilla's fidelity for the full $|\alpha_{Steane}\rangle = (|0_L\rangle + |1_L\rangle)/2^{1/2}$ state using Network-1, vs. (a) Evolution error $\varepsilon$, (b) gate error $\gamma/7$. The lines appear in two groups depending on the $\gamma$ or $\varepsilon$ values (as printed on the figure), and represent different possibilities: (●) noisy, (O) without verification step, (✚) perfect verification step and the dashed lines without symbols are the theoretical estimation without verification.

FIG. 5: Results for the $|0_L\rangle$ state synthesis vs. $\varepsilon$ and $\gamma$, using Network-1: (●) with noisy verification, (✚) perfect verification and (O) without verification. Continuous lines ($\varepsilon$ or $\gamma = 0.001$), dashed lines ($\varepsilon$ or $\gamma = 0.01$).

FIG. 6: Results for the noisy ancilla state synthesis vs. $\varepsilon$ and $\gamma$: (●) Network-1, (O) Network-3 and (★) Shor's cat state. Continuous lines ($\varepsilon$ or $\gamma = 0.001$), dashed lines ($\varepsilon$ or $\gamma = 0.01$).

FIG. 7: Network-3

FIG. 8: Information-qubit syndrome measurement network, using two ancilla states $a_z$ and $a_x$.

FIG. 9: Some error paths infecting the information-qubit with: (a) Phase errors and (b) bit-flip errors. All phase errors are represented as "bangs" and bit-flips as "X".

FIG. 10: Whole recovery network scheme for a three syndrome Steane's method and Shor's one syndrome measurement.

FIG. 11: Information-qubit recovery fidelity vs. evolution and gate error. (a) Upper group of lines are for $\gamma=0.01$, lower $\gamma=0.001$. (b) Upper group $\varepsilon=0.01$, lower $\varepsilon=0.001$. Dashed lines and unfilled symbols are one-syndrome correction and solid lines and filled symbols are three-syndromes correction: (▲,△ Steane's Network-1), (◆,◇ Network-3) and (■,□ Shor).

FIG. 12: Information-qubit recovery fidelity vs. evolution error $\varepsilon$ (the lower lines with fixed $\gamma=0.001$) or gate error $\gamma/7$ (the upper lines with fixed $\varepsilon=0.001$). Dashed lines with (+) are Network-3 without ancilla verification. Solid lines with (◆) are Network-3 with ancilla verification.

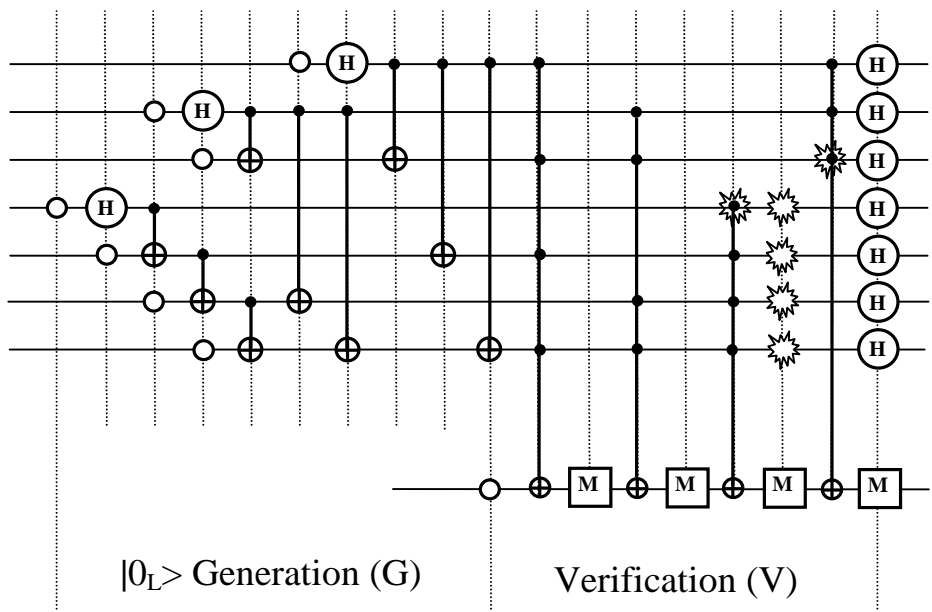

**Fig. 1 (P.J. SALAS)**

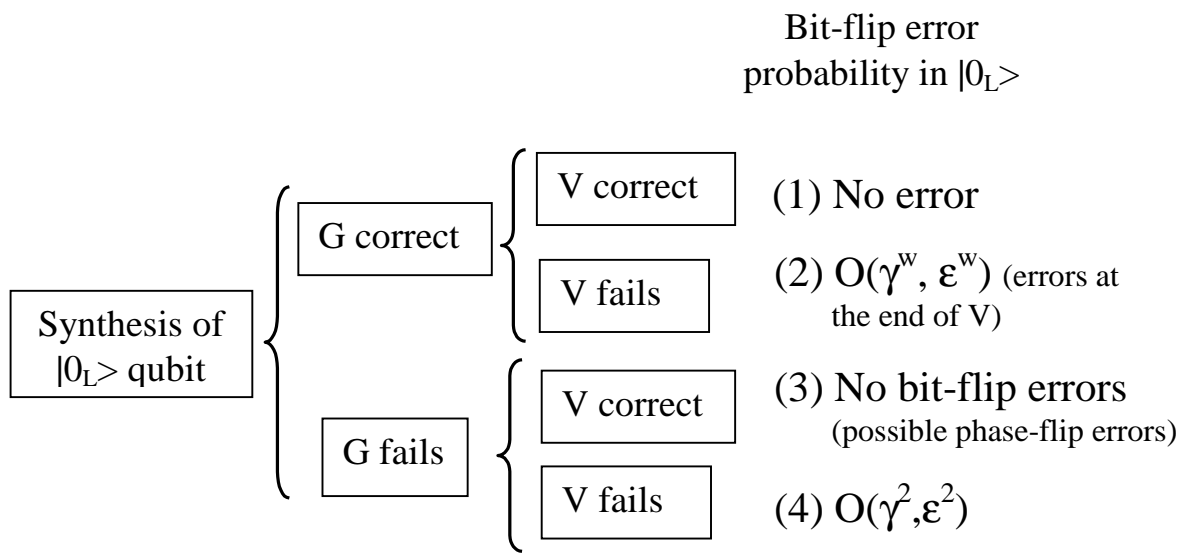

**Fig. 2 (P.J. SALAS)**

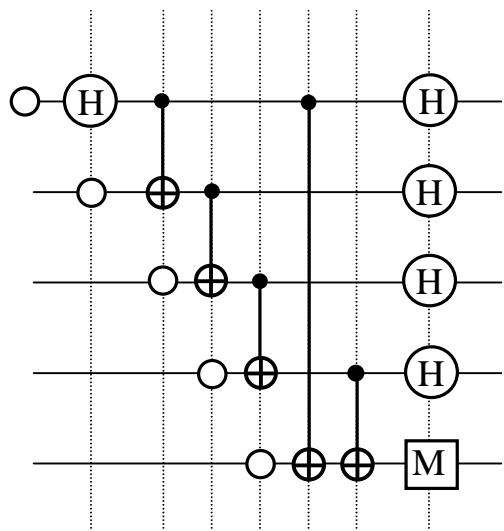

**Fig. 3 (P.J. SALAS)**

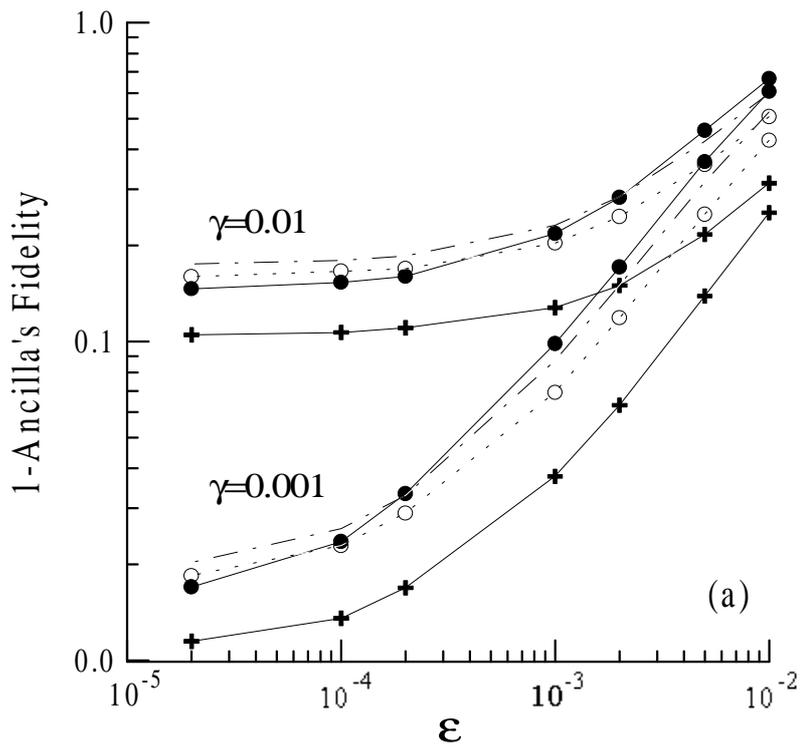

Fig. 4(a) (P.J.Salas)

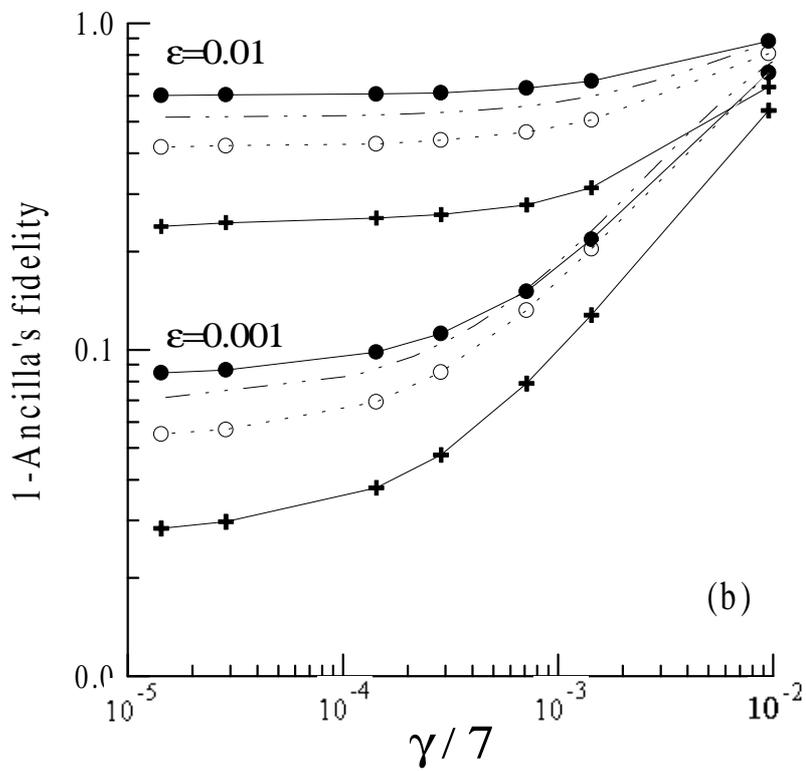

Fig. 4(b) (P.J.Salas)

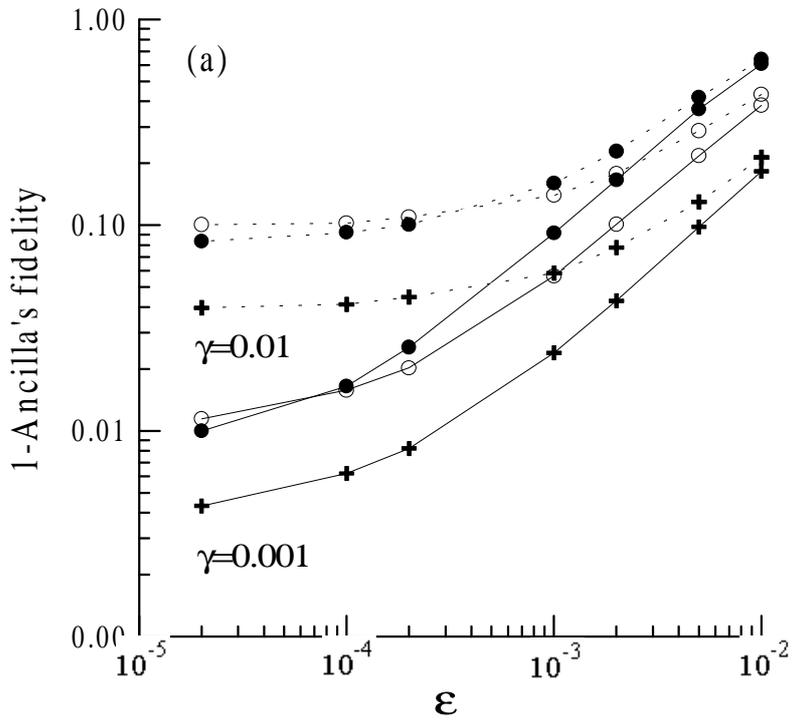

Fig. 5(a) (P.J.Salas)

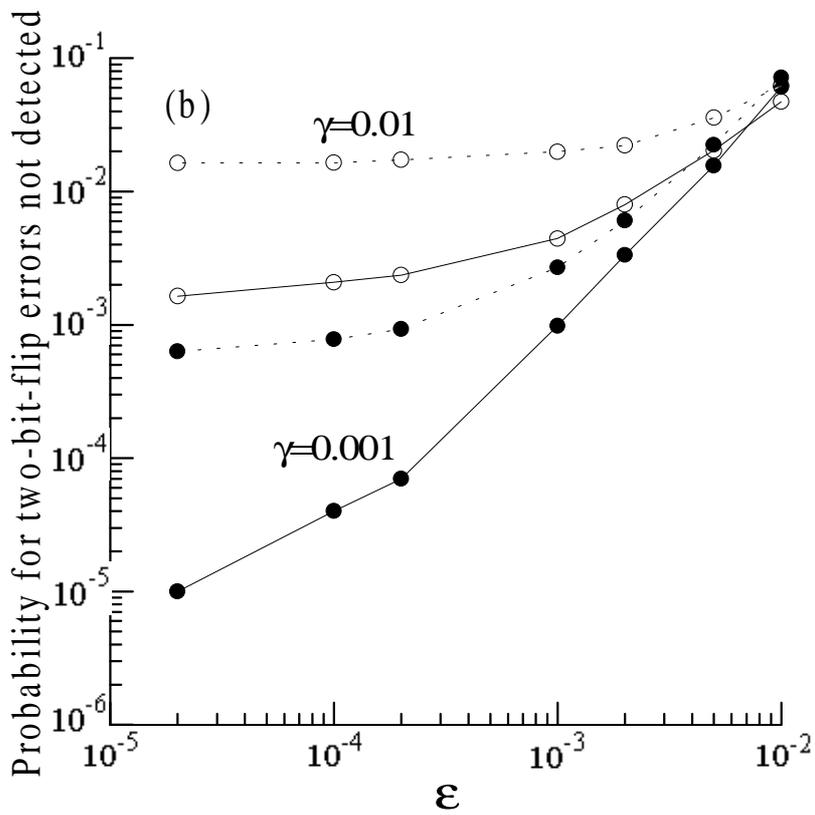

Fig. 5(b) (P.J.Salas)

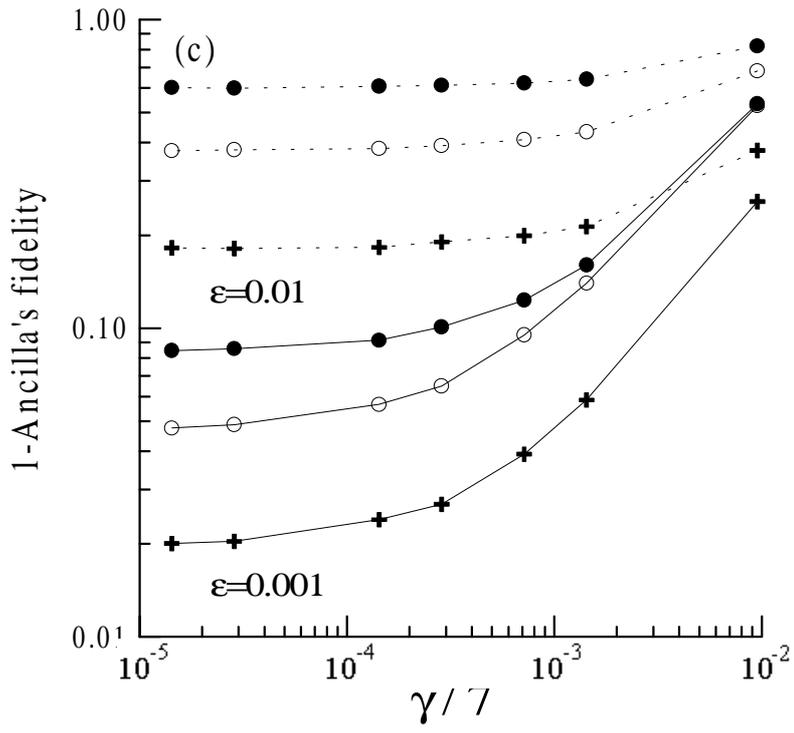

Fig. 5(c) (P.J.Salas)

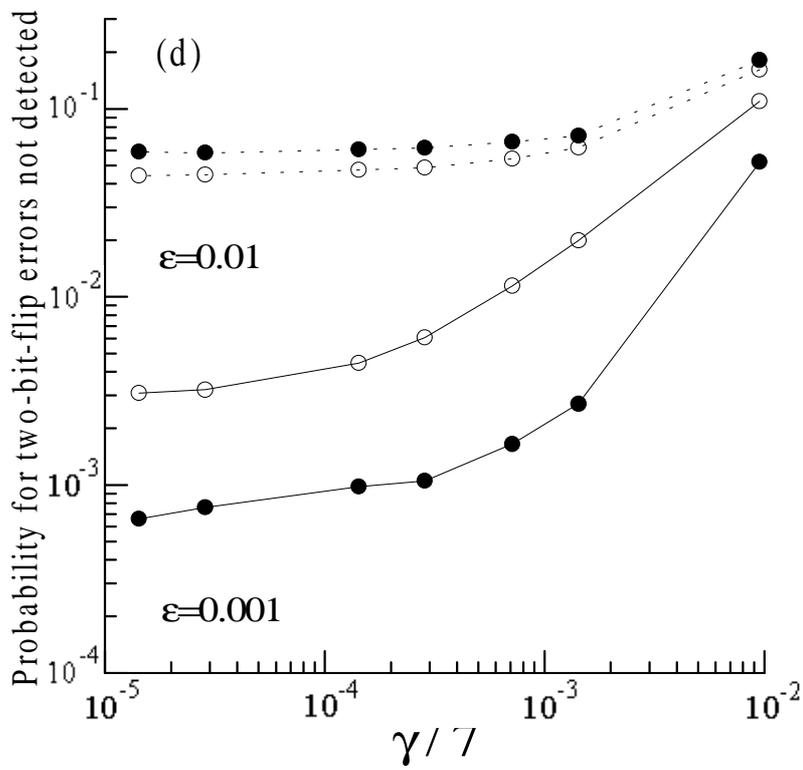

Fig. 5(d) (P.J.Salas)

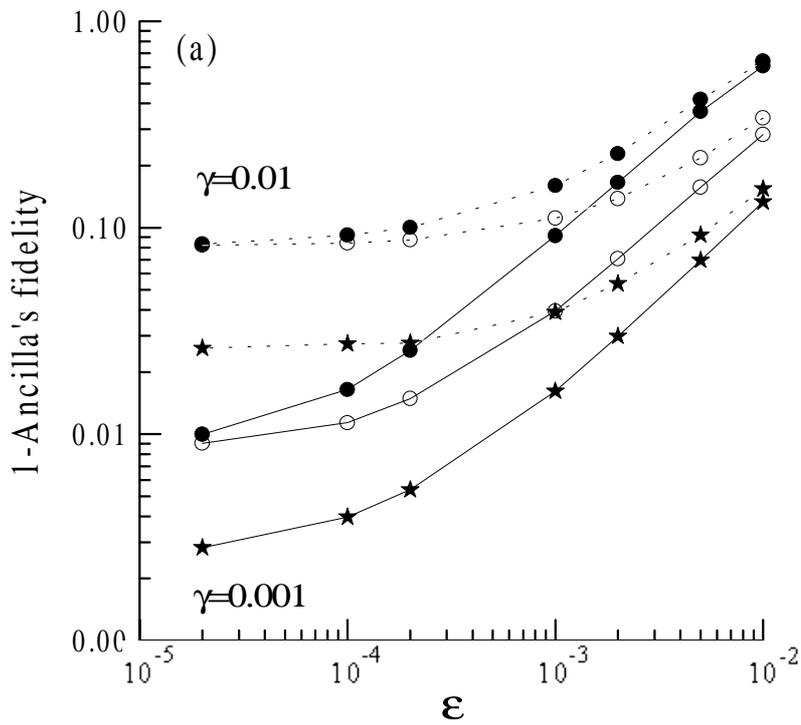

Fig. 6(a) (P.J.Salas)

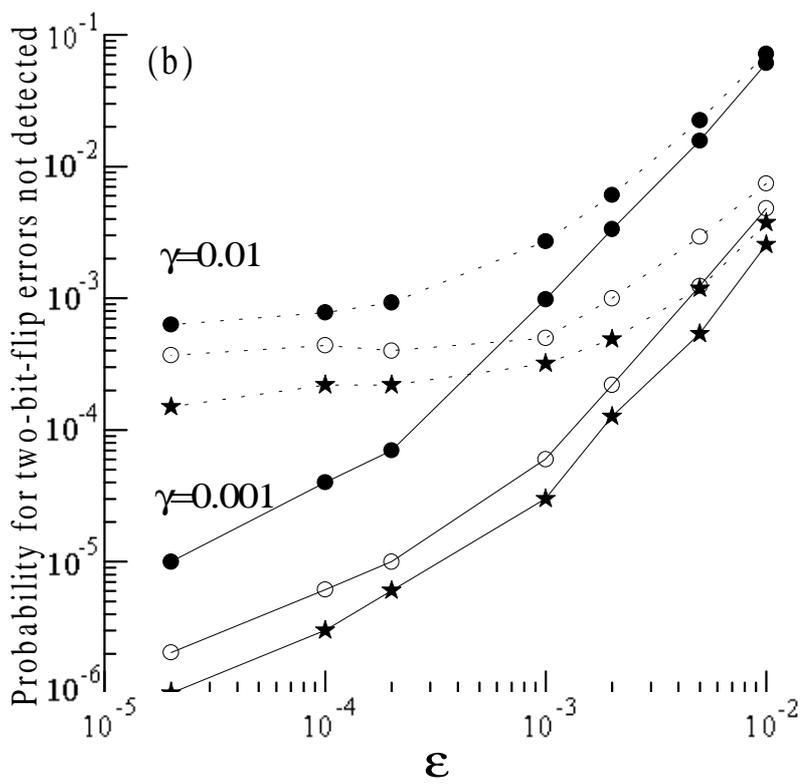

Fig.6(b) (P.J.Salas)

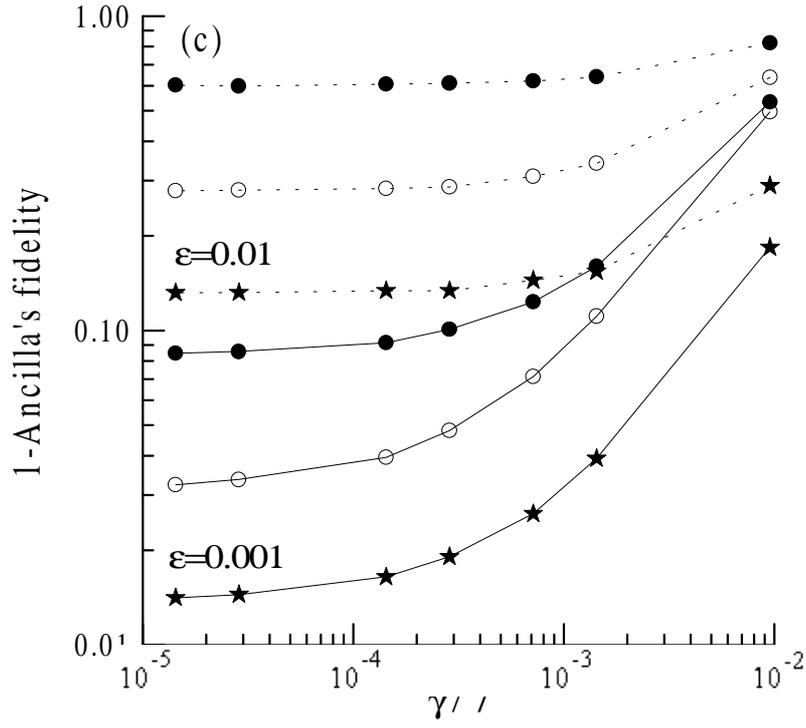

Fig. 6(c) (P.J.Salas)

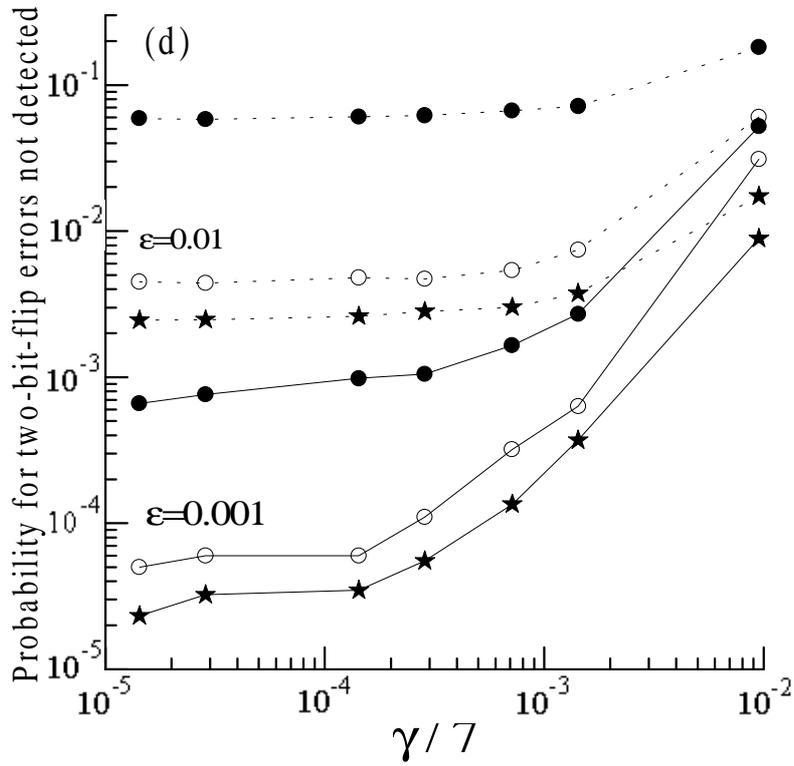

Fig. 6(d) (P.J.Salas)

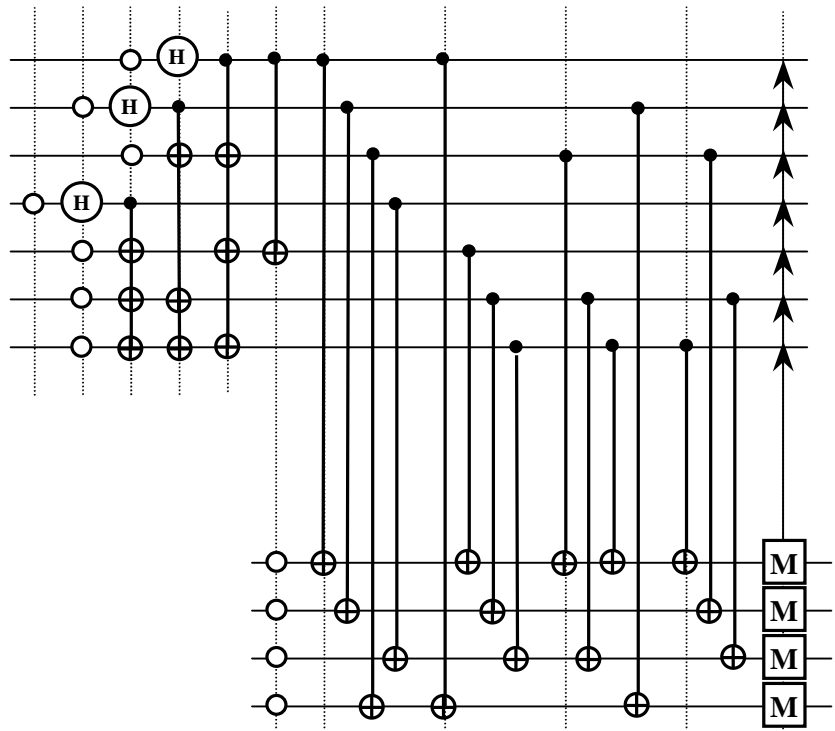

**Fig. 7 (P.J. SALAS)**

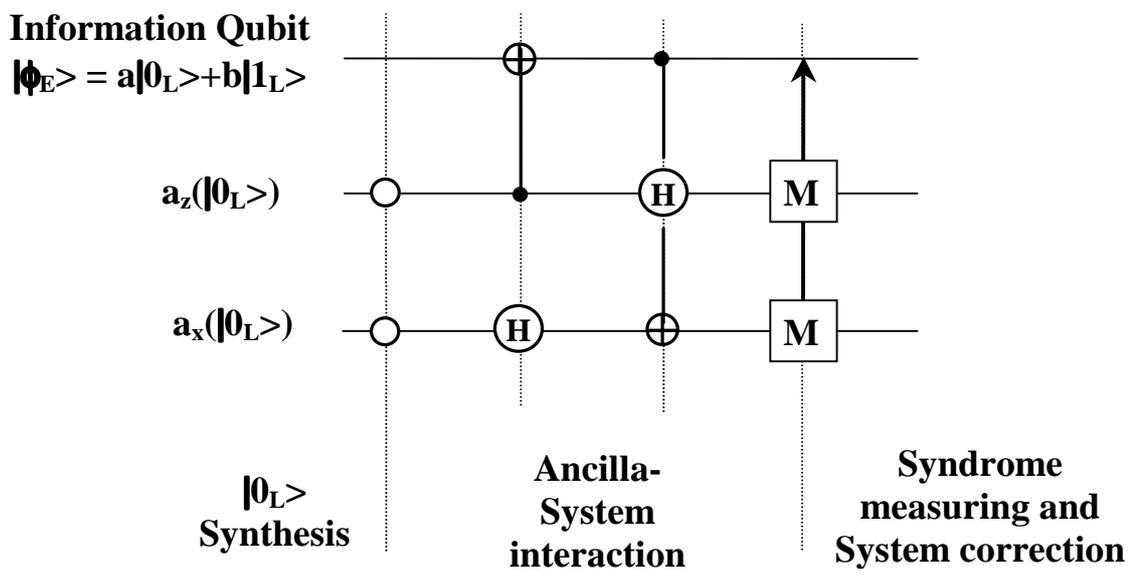

Fig. 8 (P.J. SALAS)

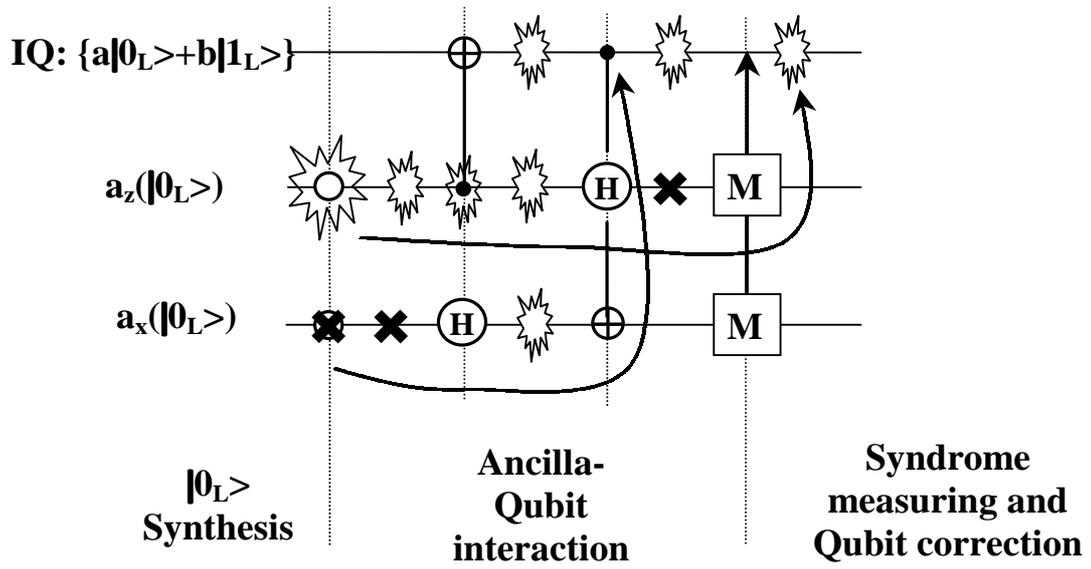

**Fig. 9(a)   (P.J. SALAS)**

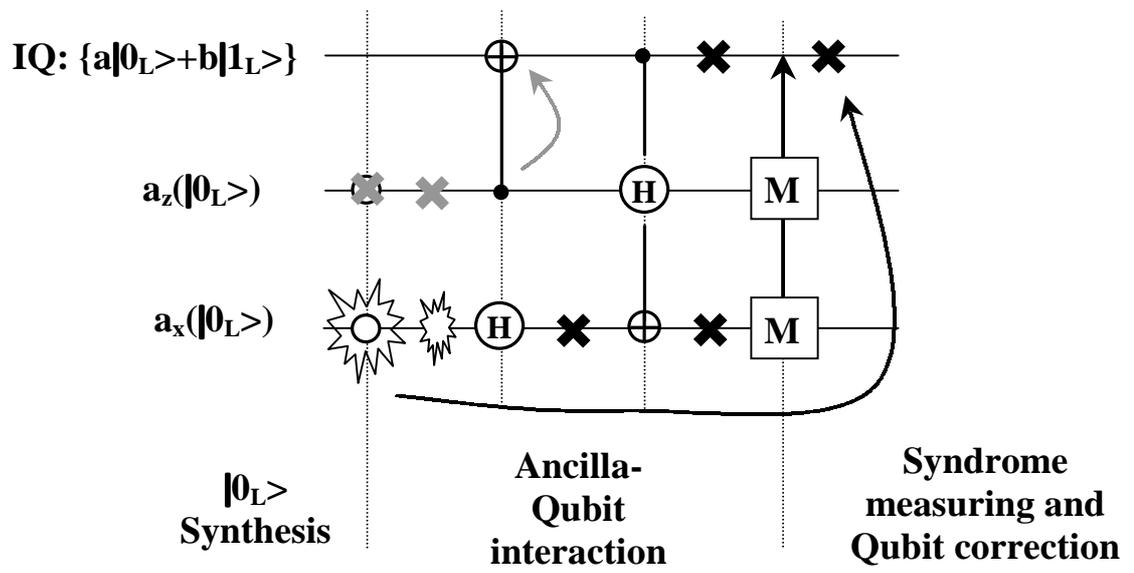

**Fig. 9(b) (P.J. SALAS)**

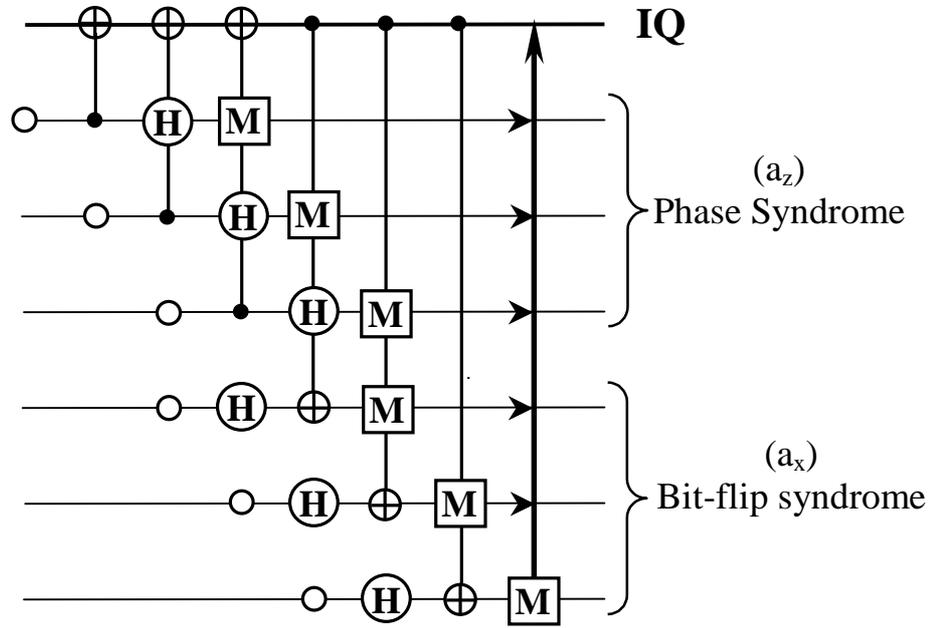

**Fig. 10** (P.J. SALAS)

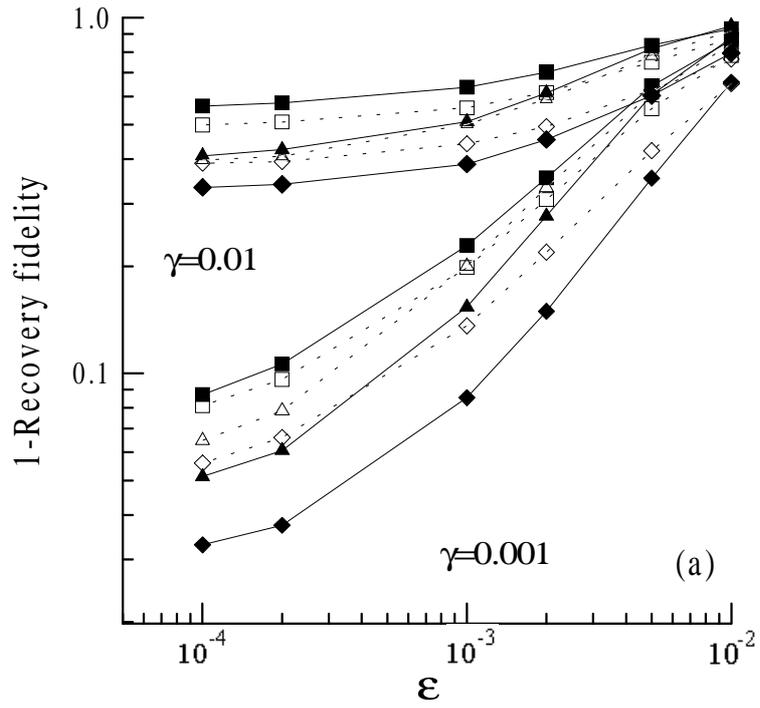

Fig. 11(a) (P.J.Salas)

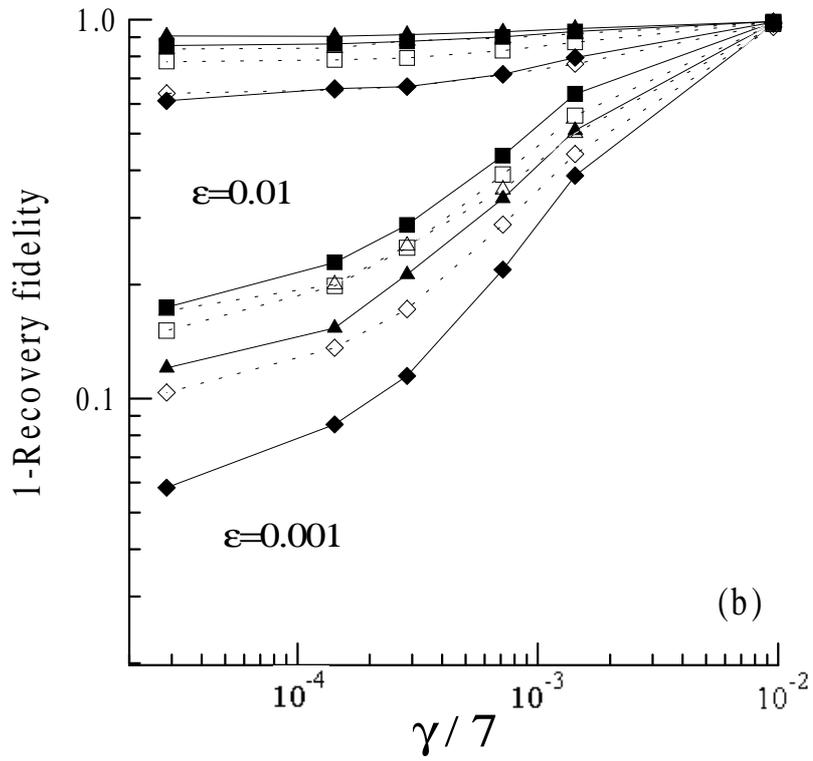

Fig. 11(b) (P.J.Salas)

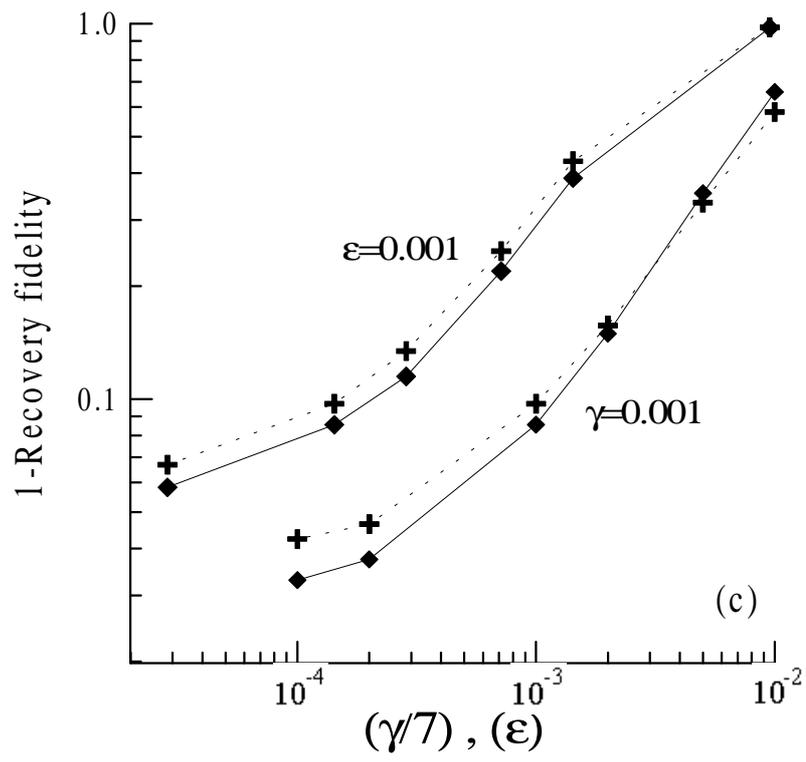

Fig.12 (P.J.Salas)